\documentclass[aps,twocolumn,showpacs,superscriptaddress,floatfix]{revtex4}
\usepackage{graphicx}
\usepackage{psfrag}
\begin{document}
\newcommand{\beq}{\begin{equation}}
\newcommand{\eeq}{\end{equation}}
\newcommand{\ben}{\begin{eqnarray}}
\newcommand{\een}{\end{eqnarray}}
\newcommand{\bea}{\begin{array}}
\newcommand{\eea}{\end{array}}
\newcommand{\om}{(\omega )}
\newcommand{\bef}{\begin{figure}}
\newcommand{\eef}{\end{figure}}
\newcommand{\leg}[1]{\caption{\protect\rm{\protect\footnotesize{#1}}}}
\newcommand{\ew}[1]{\langle{#1}\rangle}
\newcommand{\be}[1]{\mid\!{#1}\!\mid}
\newcommand{\no}{\nonumber}
\newcommand{\etal}{{\em et~al }}
\newcommand{\geff}{g_{\mbox{\it{\scriptsize{eff}}}}}
\newcommand{\da}[1]{{#1}^\dagger}
\newcommand{\cf}{{\it cf.\/}\ }
\newcommand{\ie}{{\it i.e.\/}\ }

\title{Quantum Signatures of The Classical Disconnection Border}

\author{F.~Borgonovi}
\affiliation{Dipartimento di
Matematica e Fisica, Universit\`a Cattolica, via Musei 41, 25121
Brescia, Italy}
\affiliation{I.N.F.M., Unit\'a di Brescia and 
I.N.F.N., Sezione di Pavia, Italy}
\author{G.~L.~Celardo}
\affiliation{Dipartimento di
Matematica e Fisica, Universit\`a Cattolica, via Musei 41, 25121
Brescia, Italy}
\affiliation{Theoretical Division and CNLS, Los Alamos National
Laboratory, Los Alamos, New Mexico 87545}
\author{G.P. Berman}
\affiliation{Theoretical Division and CNLS, Los Alamos National
Laboratory, Los Alamos, New Mexico 87545}

\begin{abstract}
A quantum Heisenberg model with anisotropic coupling and all-to-all
interaction  has been  analyzed  using the Bose-Einstein statistics.
In  Ref.\cite{jsp} the existence 
of a classical energy disconnection border (EDB) in the same kind of models 
has been demonstrated.
We address here the problem to find  quantum signatures of the EDB.      
An independent definition 
of a quantum disconnection border, motivated by 
considerations strictly valid in the quantum regime is given.
We  also discuss the  dynamical relevance of the quantum border  
with respect to quantum magnetic reversal.
Contrary to the classical case the magnetization can flip
even below the EDB through Macroscopic Quantum Tunneling. 
We evaluate the time scale for magnetic reversal
from statistical and spectral properties, 
for a small number of particles and 
in the semiclassical limit. 
\end{abstract}

\date{\today}
\maketitle

The existence 
of an energy disconnection border (EDB)
(previously called non-ergodicity threshold),
in classical Heisenberg models
with anisotropic  coupling and infinite range of  interaction  
has been recently found\cite{jsp}.
Below the EDB 
the energy surface is disconnected into two components
with opposite sign of the total magnetization.
The dynamical  consequences of the EDB in 
classical Heisenberg models with infinite range coupling
has also been investigated \cite{firenze}.
In particular, it has been shown that below the EDB 
the magnetization cannot change sign in time,
while above it,  in a fully
chaotic regime, a time scale for the magnetic reversal
(time needed for the total magnetization to flip)
can be determined.
Magnetic reversal times show an exponential growth with the number of 
spins and, as in standard phase transitions, 
a power law divergence at the EDB itself.

The existence of this border is not limited 
to the infinite range coupling case and can be in general 
related to the anisotropy of the
coupling when it induces an easy--axis of magnetization
(defined by the direction of the magnetization
in the minimal energy configuration of the system).
The relation between the EDB and the range of the interaction has also been
studied: 
we have taken 
into account a Heisenberg models with
an interaction potential 
among the spins which decays as  $R^{-\alpha}$,
where $R$ is the distance among the spins.
Defining as  $r$, the ratio of the disconnected portion
of the energy range with respect to the total energy range,
it has been  proved that for a $d-$dimensional system 
$r$ tends to zero in the thermodynamic limit for 
$\alpha>d$(short range) while it remains finite
for $\alpha<d$ (long range) \cite{brescia}. 

The  results found in the classical model guided our investigations 
on  the quantum side.
We are mainly interested here in the quantum signature of the
classical EDB, and on its
relevance with respect to the quantum reversal time
of the magnetic moment.

We consider here an  infinite-range 
interacting system since the explicit expression
of the EDB  has been obtained 
in this case only.
Despite its unphysical character, magnetic 
systems  can be realized,
within modern experimental techniques \cite{cornell}, described by 
Heisenberg--like  Hamiltonians with an infinite range term,
which could induce the presence of the EDB.
Moreover, when the range of the interaction is of
the same order of the size of the system,
the all-to-all coupling could be an important first order approximation
in the understanding of their behavior~\cite{B1,thebibble}.
This could be the case for small systems used in 
present nano-technology which requires to deal with 
systems with a few dozens of particles\cite{Nat1}, 
or for macroscopic systems with long range interactions.

We  first analyze the spectral properties  
and we establish the existence                      
of a quantum disconnection border in close correspondence     
with the classical one. 
An analytical estimate of this quantum threshold is given. 
We will then study the system from a dynamical point of 
view, analyzing  the time scale for quantum magnetic reversal and 
comparing  the quantum magnetic reversal times with the classical   
ones.

We consider a system of $N$ particles of spin $l$,
described by the following Hamiltonian,
\begin{equation}
\label{eq:quant_ham}
\hat {H} =\frac{\eta}{2} \sum_{i=1}^N \sum_{j\ne i} \hat{S}_i^x 
\hat{S}_j^x 
-\frac{1}{2} \sum_{i=1}^N \sum_{j\ne i} \hat{S}_i^y \hat{S}_j^y,
\end{equation}
where $-1 < \eta \le 1 $ is the anisotropy constant.
We define $\hat{M}_{x,y,z}=\sum_i S_i^{x,y,z}$ as the components
of the total magnetization of the system.
Due to the anisotropy of the coupling the system 
has an easy--axis of the magnetization along the $y-$direction.
Quantization of the Hamiltonian follows the standard rules.
According to the correspondence principle, the
classical limit is recovered as $l \rightarrow \infty$.
As in the classical case we fix the modulus
of the spins to one. This can be achieved with
an appropriate rescaling of the 
Planck constant,  $\hbar \to \hbar /|S_i|=1/\sqrt{l(l+1)}$.
With this choice, in the classical limit, $l \rightarrow \infty$
($\hbar \to 0$),  
the spin modulus remains equal to $1$.
Because of the infinite range nature of the interaction,
the Hamiltonian (\ref{eq:quant_ham}) is a completely 
symmetric operator with  respect to particle exchange.
It is thus natural to limit ourselves
to subspaces of definite symmetry.
Specifically,  we  consider the bosonic case 
(an ensemble of integer spins),
so we will limit our analysis in the subspace of 
all possible completely
symmetric states, with dimension ${\cal N} = (N+2l)!/(N! (2l)!)$. 
This choice reduces  considerably the dimension of 
the Hilbert space, allowing to extend our analysis
further in the classical limit.
An important property of the Hamiltonian (\ref{eq:quant_ham}) is  
its invariance under a $\pi$ rotation 
about  the $z$-axis :
the Hamiltonian commutes with the operator
$\exp (i \pi \sum \hat{S}_i^z )$, and  its eigenstates can be labeled
as odd (-) or even (+)  according to whether they  
change or do not change  sign
under such rotation.

\begin{figure}
\includegraphics[scale=0.3]{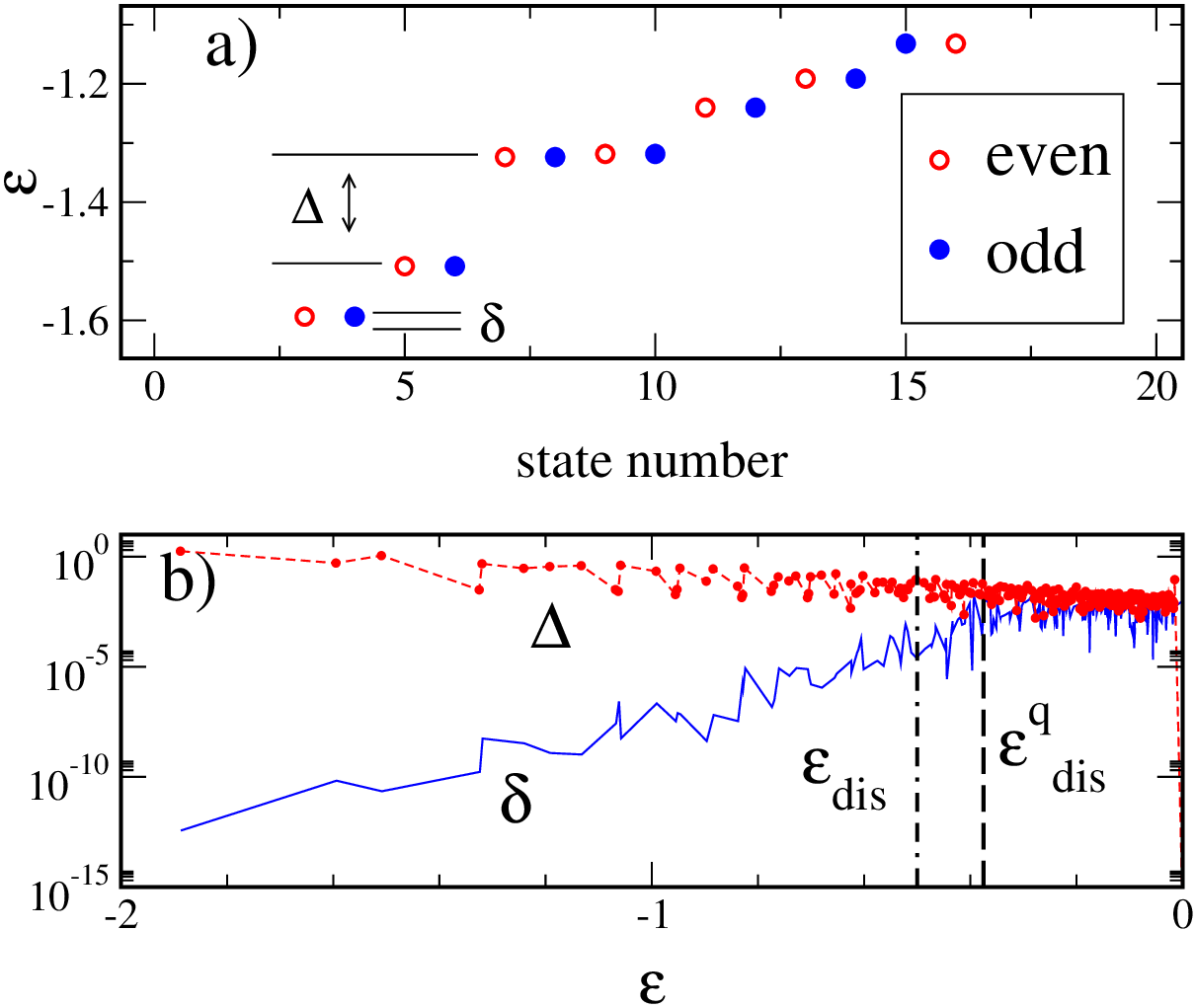}
\caption{(Color online) $N=6$ $l=3$  $\eta=1$.
a)  Doublet structure of the low energy region of the spectrum.
The different parity of the states constituting the doublets is
shown. We also indicated the level distance among even eigenstates, $\Delta$.
b) the splitting of the doublets, $\delta$,  {\it vs}
the specific energy, $\epsilon= E/N$ (blu line) and
the nearest neighbor level spacing $\Delta(\epsilon)$ (red line).
Also shown as vertical lines:
the quantum disconnection border, $\epsilon_{dis}^q$ (dashed),
 computed analytically from Eq.(\ref{eq:eqth}),
and $\epsilon_{dis} $ (dashed-dotted).
}
\label{ne1}
\end{figure}

The first aim of our analysis on the quantum system
is to assess the quantum signature of 
the classical disconnection threshold, $\epsilon_{dis}$.
Numerical diagonalization of  (\ref{eq:quant_ham}) gives 
rise to a quasi-degenerate energy spectrum 
with a energy splitting increasing with the energy.
It is a standard result\cite{kubo},
that the infinite time average of any quantum operator is zero
in presence of a non-degenerate discrete spectrum
and if its quantum average over the energy eigenstates is zero.
The operator $\hat{M}_y$  satisfies
this conditions 
thus the total magnetization along the easy--axis, 
can change its sign for any energy. 
Also, the time scale  at which this happens can be obtained by a
detailed study of the energy difference between close eigenstates.
Specifically, since the matrix elements of  
$\hat{M}_y$ between
energy eigenstates of the same parity is zero, it will be important to study
the characteristics of the energy distance between even and odd eigenstates.
The possibility for the magnetization to reverse its sign 
also in the energy region where it would be classically forbidden,
can be interpreted as a manifestation of  
Macroscopic Quantum Tunneling \cite{chud}, 
(the total magnetization can be  a macroscopic quantity).
The evaluation of the tunneling rates becomes then 
crucial to obtain the time scale for the quantum magnetic reversal.

The most evident property of the energy spectrum is 
the presence, in the low energy region, of quasi degenerate
doublets, see Fig.~\ref{ne1}a. 
Each doublet is composed by an even and an odd eigenstate.
Even if from Fig.~\ref{ne1}a they seem degenerate, they are
actually split by a small  energy difference $\delta$.
At high energy the doublets are not well defined anymore.
In Fig.~\ref{ne1}a we have also indicated the level spacing
between neighbor  even states $\Delta$. 
Note that the doublets are well defined only when  $\delta  \ll \Delta$.

\begin{figure}
\includegraphics[scale=0.3]{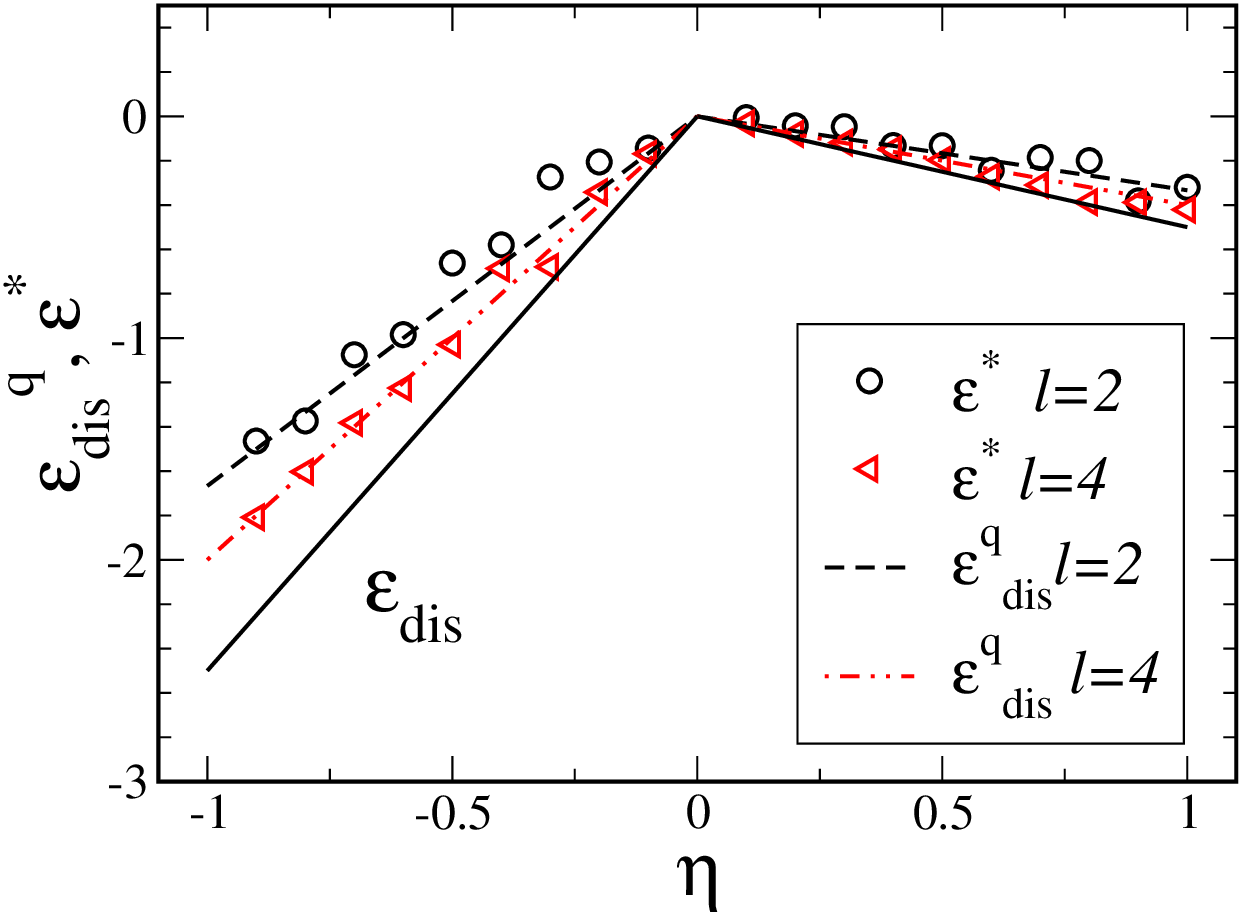}
\caption{(Color online)
$\epsilon_{dis}^q$ (dashed lines) and
$\epsilon^*$ (symbols) for different  $l$ values
{\it vs} $\eta$.
Data are  $N=6$, $l=2,4$.
Also shown (as full line)  $\epsilon_{dis}$ that
showing, at the same time,
the agreement between $\epsilon_{dis}^q$ and $\epsilon^*$
and their common semiclassical limit $\epsilon_{dis}$.
}
\label{eqth}
\end{figure}

The splittings of the doublets,  $\delta$, can be computed
numerically  for each even (or odd) eigenstate 
taking  the energy difference from its closest  odd (or even)
state;
$\delta$ can thus be considered as a degree of the spectrum degeneracy.
In Fig.~\ref{ne1}b we show how $\delta$ varies with the specific energy,
$\epsilon=E/N$: a change of slope is clearly visible, from exponential
to almost constant.
Note that the energy at which the slope changes is 
close to the classical $\epsilon_{dis}$, see Fig.~\ref{ne1}b .
It is possible to 
have a better estimate of the energy at which
the slope changes, considering the energy  $\epsilon^*$
at  which the doublets disappear:
$\delta(\epsilon^*) \simeq \Delta(\epsilon^*)$, see Fig~\ref{ne1}b.
Note that the value $\epsilon^*$,
distinguishes two regions of the spectrum,
characterized or not by  quasi-degenerate
eigenvalues pairs.
The behavior of $\epsilon^*$ for different  $l$  values
as a function of the anisotropic coefficient $\eta$
has been reported in Fig.~\ref{eqth}.
As one can see $\epsilon^* \rightarrow \epsilon_{dis}$ as
$l$ increases, which confirms
the soundness of our definition.

$\epsilon^*$ can be thought 
as a quantum correction to the classical EDB.
An hand-waving argument allows to  evaluate this quantum correction:
in the classical case, $\epsilon_{dis}$ has been obtained
computing the minimum of 
$-(\eta/2)  \sum (S_i^x)^2$
when $\eta > 0$, and  of 
$(\eta/2)  M_x^2 - (\eta/2)  \sum (S_i^x)^2$
when $\eta < 0$\cite{jsp}.
Thus, the lowest eigenvalue of the same
operators  
could give an approximate quantum border.
We will call this threshold the quantum disconnection border
$\epsilon_{dis}^q$, and we have:

\begin{eqnarray}
\label{eq:eqth}
\epsilon_{dis}^q &\sim &-\frac{\eta}{2} 
 (\hbar l)^2 \hspace{0.2cm} {\rm for}\hspace{0.2cm} 
\eta >0 \nonumber \\
\epsilon_{dis}^q &\sim& \frac{\eta}{2}  (N-1) 
(\hbar l)^2 \hspace{0.2cm} {\rm for } \hspace{0.2cm} \eta <0.
\end{eqnarray}

The agreement between the numerical values $\epsilon^*$
and our analytical estimate  $\epsilon_{dis}^q$
has been shown  in Fig.~\ref{eqth}
(compare symbols with dashed and dotted lines).

\begin{figure}
\includegraphics[scale=0.3]{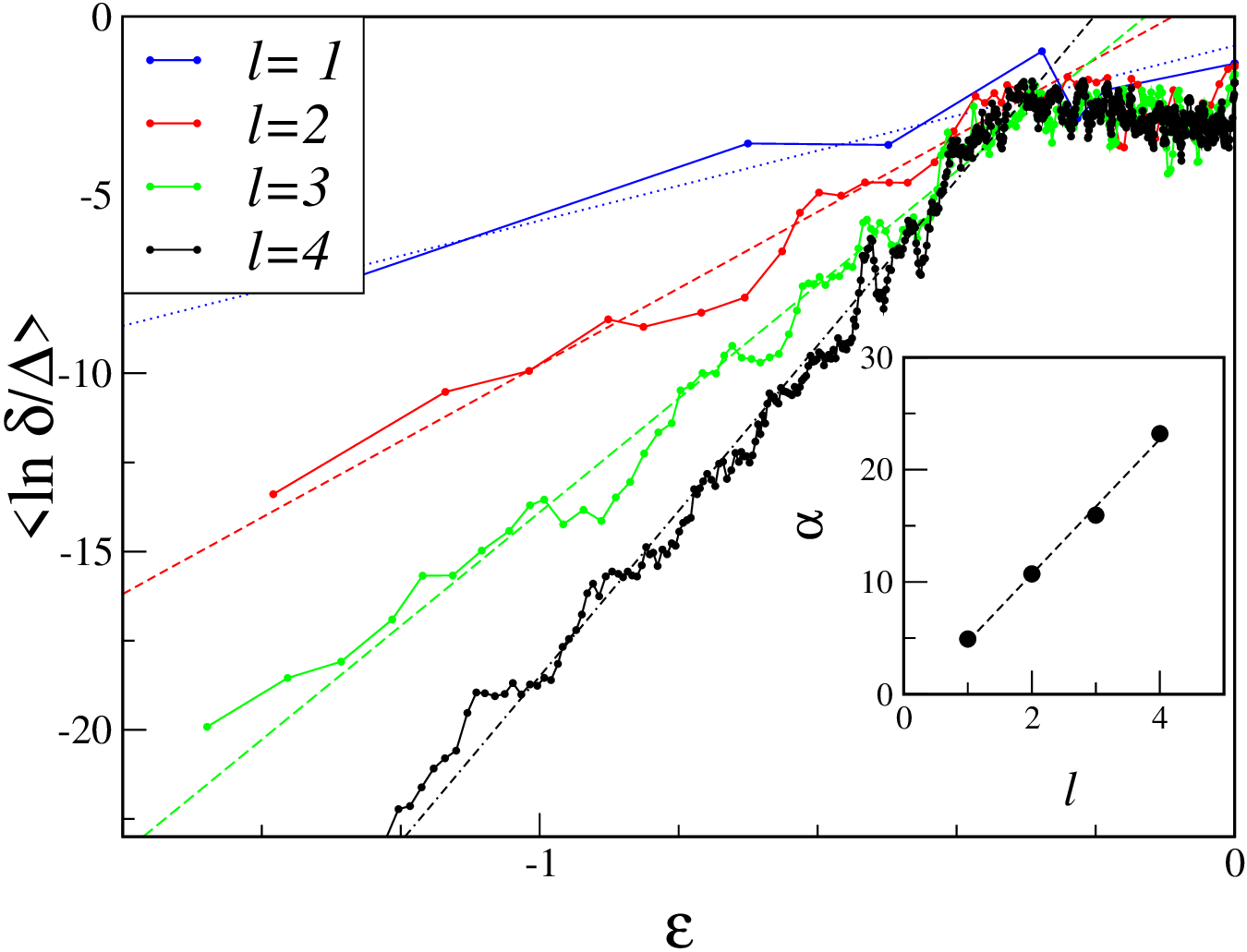}
\caption{(Color online) Average (over energy bins)
of $\langle \ln (\delta/D) \rangle$ {\it vs} 
the specific energy $\epsilon$,
for $N=6$, $\eta=1$  and different $l$, as indicated in the legend.
Here $D$ is the average level spacing.
Also shown (dashed and dotted style)  the linear fit.
Inset : the linear dependence of the slopes $\alpha$
with $l$. Linear fitting is $\alpha =  -1.3+ 6 l $. }
\label{dep}
\end{figure}

As shown in Fig.~\ref{dep} the  level splittings increase exponentially
with the energy. Also, on increasing the semiclassical parameter
$l$, the rate of growth becomes steepest.
In Fig.~\ref{dep} we show the average of $\ln \delta$ 
over suitable energy bins,
and normalized to the average level spacing $D$ 
(obtained dividing the energy range by the number of states).
Linear fits have also been indicated as dashed lines.
In the inset the exponent $\alpha$, obtained from the fitting :
$\langle \ln (\delta /D) \rangle = \alpha \epsilon  + C$
 has been shown to have a linear
dependence on the semiclassical parameter $l$, so that 
$\delta(E) \approx  0$ for $\epsilon < \epsilon_{dis}$ when $l\to\infty$.

Let us now analyze the time scale for magnetic reversal
in the quantum system, comparing the results with the classical 
ones\cite{firenze}.
Since  in the classical case the reversal times have been determined
at a fixed  energy (microcanical approach), we adopt here 
the same procedure and compute the reversal time of
the quantum average magnetization
starting from 
an ensemble of initial states, $| \psi \rangle$, obtained choosing randomly  
energy eigenstates in a narrow  energy interval:
$ |\psi \rangle = \sum_E^{E+\Delta E}  C_E |E \rangle$. 
The coefficients $C_E$ have been randomly chosen in modulus and phase
and such that $\sum_E^{E+\Delta E}  |C_E|^2 = 1$.
Since the total magnetization along the easy--axis,
$\hat{M}_y$, connects only energy 
eigenstates with different parity, we have:
\begin{equation}
\label{eq:dyn}
\nonumber \langle M_y(t) \rangle = 2 
{\cal R}e \{ \sum_{E_+,E_-=E}^{E+\Delta E} C^*_{E_+} C_{E_-}
  e^{-it/ T}
 \langle E_+| \hat{M}_y| E_-  \rangle \},
\end{equation}              
where $T=\hbar/ (E_--E_+)$.
>From $\langle M_y(t) \rangle$ we compute 
the time of first passage through zero
for each initial state of the ensemble.
>From these times we obtain 
the average magnetic reversal time $\tau$.
Before presenting the results of our analysis let us recall 
that in the quantum case, at variance with the classical one,
we are legitimate to ask what is the time scale for magnetic reversal
in the whole energy range.
Indeed, since  the
energy spectrum is  non-degenerate, from Eq.(\ref{eq:dyn}),
the average magnetization will soon or later reverse its sign,
even below the EDB.

\begin{figure}
\includegraphics[scale=0.28]{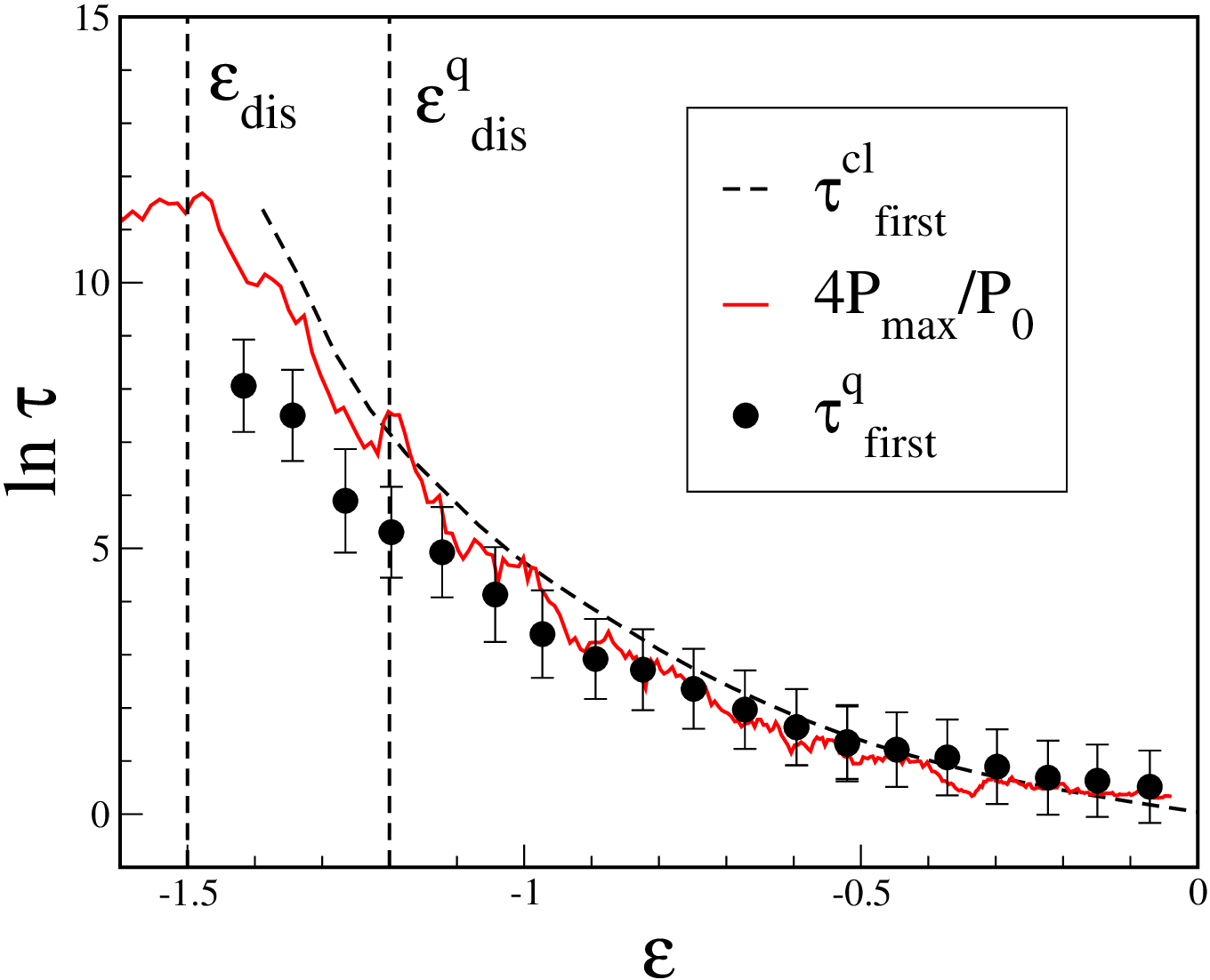}
\caption{ (Color online)
The quantum average reversal time, $\tau$ (circles),
as a function of $\epsilon$ is are shown  for the case
$N=6$, $\eta=1$ and $l=4$, are compared with the classical ones
(dashed black line),
showing a good agreement above $\epsilon_{dis}^q$ and a
 deviation near $\epsilon_{dis}$.
Also shown, as a full line $4 P_{max}/P_0$
averaged over close eigenfunctions.
}
\label{fig:qtimes}
\end{figure}

\begin{figure}[t]
\includegraphics[scale=0.3]{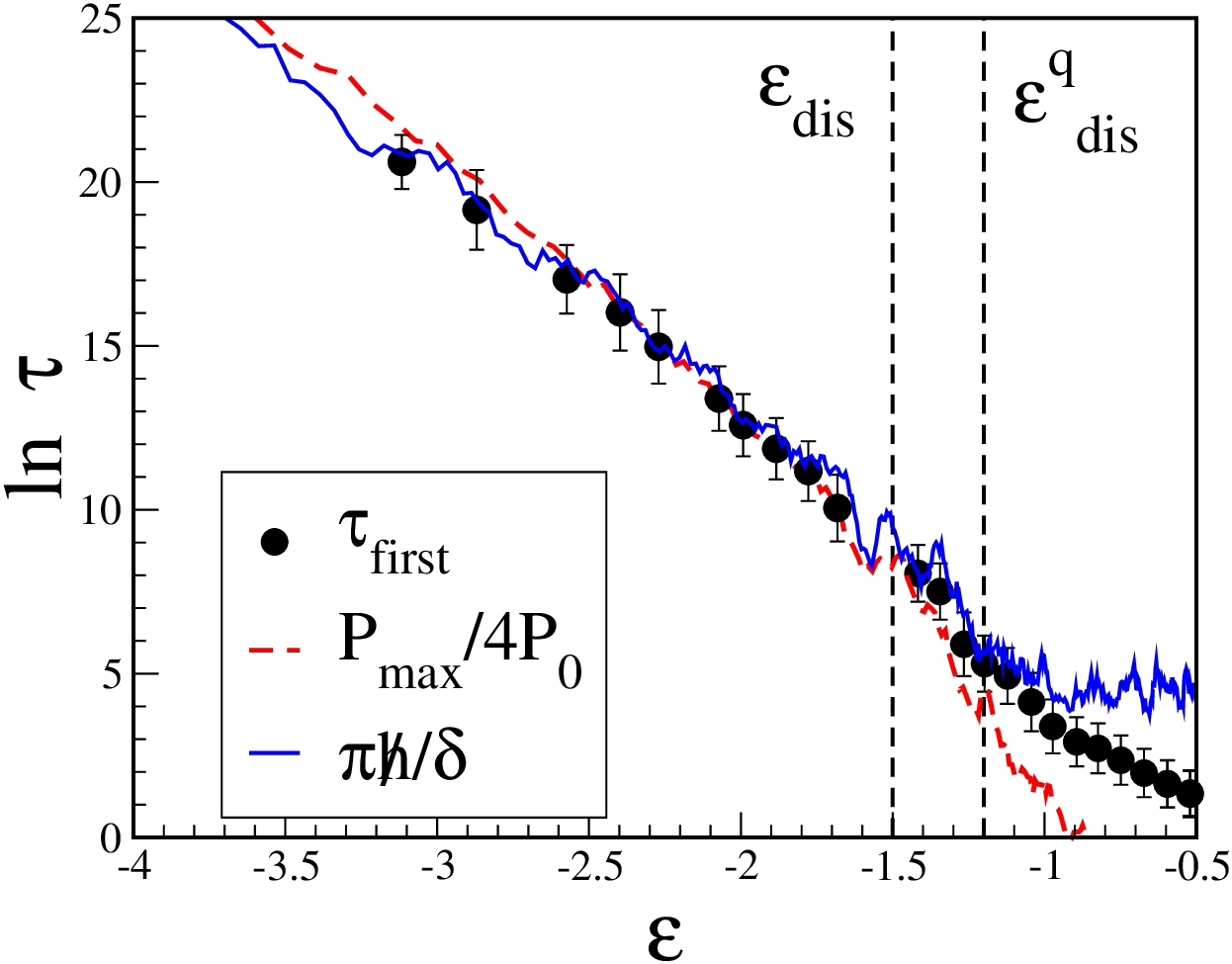}
\caption{(Color online).
Comparison between dynamical and tunneling splittings
in the classically forbidden energy region.
Circles : dynamical quantum reversal times;
Full (blu) line:  tunneling splittings
$\tau \sim  \pi \hbar /(2 \delta)$
for the case: $N=6$, $l=4$.
Not surprisingly, the  evaluation of $\tau$ through $\delta$
does not work above $\epsilon_{dis}^q$, where the doublet structure
disappear.
As  dashed (red) line $P_{max}/4P_0$.
}
\label{fig:tsotto}
\end{figure}

In Fig.~\ref{fig:qtimes} we consider the energy region
above $\epsilon_{dis}$. As one can see there is good agreement
between classical and quantum times above 
$\epsilon_{dis}^q$. 
Note that the classical times diverge at $\epsilon_{dis}$
at variance with the quantum ones which are systematically smaller,
in the region betwen $\epsilon_{dis}$ and
$\epsilon_{dis}^q$ .
This is not surprising since the 
possibility of tunneling will
enhance the probability for the 
magnetization to reverse its sign.
In the classical 
case we successfully evaluated the 
reversal times from the entropic barrier, $\Delta S$, between the most 
probable value of the magnetization and its  zero value, 
$\tau \sim e^{-\Delta S}=P_{max}/P_0$
\cite{firenze}.
Following the classical case we also computed
$P_{max}/P_0$ from
the quantum probability distribution of the magnetization, $P(M_y)$,
and compared the dynamical times $\tau$ 
with the probabilistic ones given by
$cP_{max}/P_0$ \cite{firenze}, where $c$ is a constant.
As one can see from Fig.~\ref{fig:qtimes}
the agreement between probabilistic and dynamical times
is fairly good   where  classical and 
quantum times agrees, while the agreement is less accurate in the
crossover region, between $\epsilon_{dis}$ and $\epsilon_{dis}^q$.
Let us now discuss, in details, the behavior of the quantum reversal
times below $\epsilon_{dis}^q$.
In the low energy region of the spectrum, 
due to the intrinsic quasi-degeneracy 
the dynamics  can be entirely characterized by the energy difference
$|E_{+}-E_{-}|=\delta$. This occurs if the
energy bin $\Delta E$ of the initial state is sufficiently
small so that one single doublet belongs to it.
The dynamics is thus oscillatory with a period  given by 
$2 \pi \hbar/\delta$.
Indeed under this condition, the magnetization oscillates coherently
between states with opposite sign, a phenomenon
known as Macroscopic Quantum Coherence\cite{takagi}.
This  period also represents, within a numerical
factor, the  time for the first passage to zero of 
$\langle M_y(t) \rangle$.
One thus can assume : $ \tau \sim \pi \hbar/ (2 \delta)$.
The agreement, over many orders of magnitude,
has been shown in Fig.~\ref{fig:tsotto}.
Also  below $\epsilon_{dis}^q$, we checked the 
proportionality of the reversal times with
$P_{max}/P_0$.
It is surprising that $P_{max}/P_0$ turns out
to be proportional
to the tunneling rates and then, 
when properly defined, 
to the reversal times, 
even  in the region classically forbidden (below $\epsilon_{dis}$)
where the only mechanism allowing the jumping of the barrier is
through Macroscopic Quantum Tunneling, see Fig.~\ref{fig:tsotto}.
This suggests that the mechanism producing this 
proportionality can also have
a non  classical origin. 
One should also note that
the constant of proportionality is  different
below $\epsilon_{dis}$ and above $\epsilon_{dis}^q$.
This explains the poor agreement between the statistical
and dynamical times in the crossover region 
($\epsilon_{dis}<\epsilon<\epsilon^q_{dis}$).

>From the results presented here 
we can 
address the problem of the  dynamical signature of the classical
Energy Disconnection Border.
In the semiclassical limit the crossover region becomes
very narrow, ($\epsilon_{dis}^q \rightarrow  \epsilon_{dis}$),
thus we can expect a crossover from of the reversal time
from a power law dependence on the energy, like in the
classical case\cite{firenze}, to and exponential
dependence on the energy.
Moreover Fig.~\ref{dep}  suggests how the classical
limit is recovered:  indeed  $\delta \approx 0 $ as
$l\to\infty$, for energies below $\epsilon_{dis}$,
which  is consistent with the fact that
the magnetization cannot reverse its sign 
below $\epsilon_{dis}$ in the classical system.

In conclusion we have found a quantum signature of the classical EDB
in the spectral properties of
the system leading to the definition of a 
quantum disconnection threshold,
$\epsilon_{dis}^q$
with the correct classical limit.
Below $\epsilon_{dis}^q$ the spectrum is characterized by the 
presence of quasi degenerate doublets, 
whose energy difference $\delta(\epsilon)$ 
depends exponentially on $\epsilon$.
The quantum reversal times of the total magnetization 
have been studied and  compared  with the classical ones
above $\epsilon_{dis}$.
We have also shown that the total magnetization can flip
in the energy region classically forbidden.
Quite surprisingly, quantum reversal times
(and thus the tunneling rates)
are still proportional to $P_{max}/P_0$ 
even below $\epsilon_{dis}$.

The existence of the 
classical EDB allows to address an energy region 
where to look for Macroscopic Quantum Phenomena,
which have recently raised much interest\cite{leggett}.
Indeed the fact that the total magnetization 
can reverse its sign even below the EDB can be seen as a manifestation of
Macroscopic Quantum Tunneling, a  well known phenomenon in micromagnetism,
also found experimentally\cite{scie,nat}.
Nevertheless Macroscopic Quantum Tunneling of magnetization arises
in literature \cite{chud,takagi},
from phenomenological Single--Spin Hamiltonians, 
where the single spin describes the
total magnetic moment of the system,
and no reference to the range of the interaction
has been explicitly pointed out.
On the other hand  we presented here a multiparticle system 
in which this  phenomenon  clearly arises,
in  connection with the existence of the disconnection border 
and  with  the long range nature of the interaction. 

G.L.Celardo acknowledges financial support from LANL and
Universit\`a Cattolica under the program {\it Foreign specialization
studies}.

\end{document}